\documentstyle[twocolumn,aps,epsf]{revtex}
\begin{document}
\draft
\twocolumn[\hsize\textwidth\columnwidth\hsize\csname
@twocolumnfalse\endcsname

%%%%%%%%%%%%%%%%%%%%% Title and Author(s) %%%%%%%%%%%%%%%%%%%%%%%%%

\title{Superconductivity in the Orbital Degenerate Model for Heavy
Fermion Systems}

\author{Tetsuya Takimoto\dag, Takashi Hotta\dag, and Kazuo Ueda\ddag\dag}

\address{\dag\ Advanced Science Research Center,
Japan Atomic Energy Research Institute, Tokai, Ibaraki 319-1195, Japan}
\address{\ddag\ Institute for Solid State Physics, University of Tokyo,
5-1-5 Kashiwa-no-ha, Kashiwa, Chiba 277-8581, Japan}

\date{\today}
\maketitle

%%%%%%%%%%%%%%%%%%%%%%% Abstract %%%%%%%%%%%%%%%%%%%%%%%%%%%%%%
\begin{abstract}
Magnetism and superconductivity of new heavy
fermion compounds CeTIn$_5$ (T=Co, Rh and Ir) are
investigated by applying fluctuation exchange approximation
to an orbital degenerate Hubbard model.
The superconducting phase with $d_{x^2-y^2}$-symmetry
is found to appear next to the antiferromagnetic phase with
increasing the orbital splitting energy.
The present theory suggests that the orbital splitting energy
plays a key role of controlling parameter for the quantum phase
transitions in the heavy fermion system.
\end{abstract}

%%%%%%%%%%%%%%%%%%%%% PACS number(s) %%%%%%%%%%%%%%%%%%%%%%%%%%%
%\pacs{PACS number:74.20.Mn, 71.27.+a, 71.10.Fd, 74.70.Tx}

\vskip2pc]
\narrowtext
%%%%%%%%%%%%%%%%%%%%%%%%%% Introduction %%%%%%%%%%%%%%%%%%%%%%%%%%%%%
%\baselineskip=30pt

\section{Introduction}

Superconductivity in strongly correlated electron systems has been
one of central issues in the research field of condensed matter physics,
since the pioneering discovery of superconductivity in CeCu$_2$Si$_2$
\cite{Steglich}.
The subsequent discovery of high temperature superconductivity in
cuprates has accelerated further investigations of this subject,
leading to unambiguous identification of the unconventional nature
of $d_{x^2-y^2}$-wave pairing in cuprates.
Recently, new heavy fermion compounds CeTIn$_5$ (T=Rh, Ir, and Co)
have been discovered \cite{Ce115}.
Among them, CeRhIn$_5$ exhibits an antiferromagnetic (AFM)
transition at a N\'eel temperature $T_{\rm N}$=3.8K and
becomes superconducting only under hydrostatic pressure
larger than 15 kbar.
On the other hand, both CeIrIn$_5$ and CeCoIn$_5$ are
superconducting at ambient pressure with transition temperatures
$T_{\rm c}$=0.4K and 2.3K, respectively.
Especially, the $T_{\rm c}$ of CeCoIn$_5$ is the highest among those
yet observed for heavy fermion superconductors \cite{Sarrao}.

Reflecting the fact that CeTIn$_5$ has the HoCoGa$_5$-type
tetragonal crystal structure,
quasi two-dimensional Fermi surfaces have been observed
in de Haas-van Alphen experiments for the compounds,
consistent with the band-structure calculation results \cite{Haga}.
Concerning the superconducting state, nuclear relaxation rate
in CeTIn$_5$ exhibits $T^3$ behavior below $T_{\rm c}$ \cite{NMR}
and thermal conductivity in CeCoIn$_5$ is found
to include a component with four-fold symmetry \cite{Izawa},
strongly suggesting $d_{x^2-y^2}$-wave pairing
in the superconducting phase of CeTIn$_5$.
Furthermore, it has been shown that in the alloy system
CeRh$_{1-x}$Ir$_{x}$In$_5$, the superconducting phase appears
in the neighborhood of the AFM phase \cite{Pagliuso}.
A natural consequence of these experimental results is that
superconductivity in CeTIn$_5$ compounds is induced
by AFM spin fluctuations,
similarly to high-$T_{\rm c}$ superconductivity.

In spite of the similarities mentioned above,
here we should emphasize several differences between
heavy fermion superconductors such as CeTIn$_5$ and
high-$T_{\rm c}$ cuprates.
First of all, relevant electrons in high-$T_{\rm c}$ superconductors
are almost itinerant $3d$-electrons,
while for heavy fermion superconductors, relevant ones are
$f$-electrons, which almost localized and their dispersion is
mainly determined by hybridization with conduction electrons.
Another important difference is concerned with the electronic states
relevant to low-energy physics.
For cuprates, it is widely recognized that a single-band model for
holes in $d_{x^2-y^2}$ orbitals on the square lattice is a good starting
point. On the other hand, for heavy fermion superconductors,
several Fermi surfaces are observed in general.
Such complex electronic states may be eventually traced back to
the orbital degeneracy and relatively weak crystalline electric field
(CEF) effect of $f$-electrons compared with the $3d$-electrons
in the CuO$_2$ plane.
It implies that construction of a realistic microscopic model is
not an easy work for heavy fermion systems.
Therefore, up to now, most of theoretical investigations
for superconductivity in heavy fermion systems have been
restricted in a phenomenological level.

In this paper, we discuss the effect of orbital fluctuations on
superconductivity based on a microscopic theory applied to
a microscopic model. In the next section,
we introduce an orbital degenerate model obtained including
important characters of CeTIn$_5$.
Then, in order to study the superconducting transition induced by
spin and/or orbital fluctuations,
we develop a strong-coupling theory using fluctuation exchange (FLEX)
approximation \cite{FLEX} in which spin and orbital fluctuations
as well as the single-particle spectrum are determined self-consistently.
Finally, we discuss experimental results for CeTIn$_5$
in the light of the present theory.

\section{Model Hamiltonian}

In order to introduce a minimal model for $f$-electron systems,
we start from the local basis for a Ce$^{3+}$ ion.
Among 14-fold degenerate $4f$-electronic states,
due to the effect of strong spin-orbit coupling,
only the $j$=5/2 sextet effectively contributes to
the low-energy excitations ($j$ is total angular momentum).
Under the cubic CEF, this sextet is further split into
$\Gamma_7$ doublet and $\Gamma_8$ quartet,
where the corresponding eigen-states are given by
$|\Gamma_{7\pm} \rangle$=
$\sqrt{1/6}|\pm 5/2\rangle$$-$$\sqrt{5/6}|\mp 3/2\rangle$, 
$|\Gamma_{8\pm}^{(1)} \rangle$=
$\sqrt{5/6}|\pm 5/2\rangle$+$\sqrt{1/6}|\mp 3/2\rangle$, 
and 
$|\Gamma_{8\pm}^{(2)}\rangle$=$|\pm 1/2\rangle$.
Here $+$ and $-$ in the subscripts denote ``pseudo-spin'' up and down
in each Kramers doublet, respectively.
For simplicity, we consider only $|\Gamma_{8} \rangle$ states,
further split into two Kramers doublets as
$|\Gamma_{8\pm}^{(1)} \rangle$ and $|\Gamma_{8\pm}^{(2)} \rangle$
under the additional tetragonal CEF. 
Note that $|\Gamma_{8\pm}^{(1)} \rangle$ and
$|\Gamma_{8\pm}^{(2)} \rangle$ belong to $\Gamma_7$
and $\Gamma_6$ irreducible representations, respectively,
in the tetragonal system.
Validity of this assumption for CeTIn$_5$ will be discussed later
in this paper, but here we stress that
the Hamiltonian constructed from $\Gamma_8$ quartet is the simplest
model including essential physics of interplay between
spin and orbital fluctuations.

In order to consider itinerant features of $4f$-electrons,
we take into account nearest-neighbor hopping of $f$-electrons
by the tight-binding method \cite{Maehira}.
It should be noted that the matrix elements of the nearest-neighbor
hopping depend on not only the orbital but also the hopping direction,
since the forms of wavefunctions of $|\Gamma_{8\pm}^{(1)} \rangle$
and $|\Gamma_{8\pm}^{(2)} \rangle$ states are different from each other. 
We can determine the hopping matrix elements by assuming that
the overlap integral through the $\sigma$-bond ($ff\sigma$) is dominant. 
Since CeTIn$_5$ has a tetragonal crystal structure
and quasi two-dimensional Fermi surfaces
have been experimentally observed \cite{Haga,Settai},
it is natural to consider the two-dimensional square lattice
composed of Ce$^{3+}$ ions.
Although the hybridization of $f$ electrons with In $5p$ electronic
states may be the main source of kinetic energy for $f$ electrons,
in the present scheme such an effect is considered as 
the effective hopping amplitude for $f$ quasi-particles,
after the $p$-electron degrees of freedom are integrated out.

By further adding the on-site Coulomb interaction terms
among $f$-electrons, the effective Hamiltonian with 
orbital degeneracy is given by
\begin{eqnarray}
  H &=& \sum_{{\bf ia}\tau\tau'\sigma}
  t^{\bf a}_{\tau\tau'} f_{{\bf i}\tau\sigma}^{\dag}
                        f_{{\bf i+a}\tau'\sigma}
  -\varepsilon \sum_{\bf i}(n_{{\bf i}1\sigma}-n_{{\bf i}2\sigma})/2 
  \nonumber \\
  &+&U \sum_{{\bf i}\tau}n_{{\bf i}\tau\uparrow}
                         n_{{\bf i}\tau\downarrow}
  +U'\sum_{{\bf i}\sigma\sigma'}n_{{\bf i}1\sigma}
                                n_{{\bf i}2\sigma'},
\end{eqnarray}
where $f_{{\bf i}\tau\sigma}$ is the annihilation operator for
an $f$-electron with pseudo-spin $\sigma$ in the $\tau$-orbital state
$\Gamma_{8}^{(\tau)}$ at site ${\bf i}$,
${\bf a}$ is the vector connecting nearest-neighbor sites,
and $n_{{\bf i}\tau\sigma}$=$f_{{\bf i}\tau\sigma}^{\dag}
f_{{\bf i}\tau\sigma}$.
The first term represents the nearest-neighbor hopping of $f$-electrons
with the amplitude $t^{\bf a}_{\tau\tau'}$ between
$\tau$ and $\tau'$ orbitals along the ${\bf a}$-direction,
given by
$t^{\bf x}_{11}$=$-\sqrt{3}t^{\bf x}_{12}$=
$-\sqrt{3}t^{\bf x}_{21}$=$3t^{\bf x}_{22}$=1 
for ${\bf a}$=${\bf x}$ and
$t^{\bf y}_{11}$=$\sqrt{3}t^{\bf y}_{12}$=
$\sqrt{3}t^{\bf y}_{21}$=$3t^{\bf y}_{22}$=1
for ${\bf a}$=${\bf y}$, respectively,
in energy units where $t^{\bf x}_{11}$=1.
The second term denotes the tetragonal CEF, leading to an
energy splitting $\varepsilon$ between the two orbitals.
In the third and fourth terms, $U$ and $U'$ are the intra- and
inter-orbital Coulomb interactions, respectively.
Due to the rotational invariance in the orbital space for
the interaction part, $U'$ should be equal to $U$, when
we ignore the Hund's rule coupling, since it is irrelevant
in the large-$U$ limit for the quarter-filling case
with one $f$-electron per site.
Thus, in this paper, we restrict ourselves to the case of $U$=$U'$.
Note also that in the quarter-filling case,
the present model is virtually reduced to the half-filled single-orbital
Hubbard model in the limit of $\varepsilon$=$\infty$.

%In reality, $U$=$U'$ may be assumed, since they originate from
%the same Coulomb interactions among $f$-orbitals in the $j$=5/2 multiplet,
%and

\section{FLEX approximation}

In our previous work, we have developed a weak-coupling theory for
superconductivity based on the same orbital degenerate model described above, 
using the static spin and orbital fluctuations obtained
within the random phase approximation (RPA) \cite{manybody}.
Various superconducting phase have been found around
varieties of ordered phases,
whose boundaries are determined by the RPA instability.
In order to develop more sophisticated theory,
we should include effects of (1) mode-mode coupling and
(2) quasi-particle damping, neglected in the previous work.
Regarding (1), the RPA does not incorporate effects of mode-mode coupling
between fluctuations, but the mode-mode coupling modifies significantly
the temperature and frequency dependences of spin and orbital fluctuations. 
Concerning (2), within the weak-coupling theory for superconductivity,
damping of quasi-particles by the scattering due to spin and orbital 
fluctuations is not taken into account,
which leads to the suppression of superconductivity.
Therefore, the weak-coupling theory generally overestimates
the region of superconducting phase.

In the present paper, we apply the fluctuation exchange (FLEX)
approximation \cite{FLEX}
to the orbital degenerate model discussed in the preceding section.
The FLEX approximation has the following two features:
(1) This is a kind of the mode-mode coupling theory,
where spin and orbital fluctuations and the spectra of
$f$-electrons are self-consistently determined through the fluctuation 
exchange self-energy.
(2) It provides the Dyson-Gorkov equation
%describing the strong-coupling theory
where the normal and anomalous self-energies are obtained
on an equal footing.
Here we emphasize that the FLEX approximation has been successful
to understand consistently the normal and superconducting states
of high-$T_{\rm c}$ cuprates, in particular with reasonable
estimation of $T_{\rm c}$ \cite{Moriya}.
Therefore, it is interesting to apply the FLEX approximation
to the orbital degenerate model to understand
its properties concerning superconductivity.

In the doubly degenerate case, the Green's functions for $f$-electrons
form a 2$\times$2 matrix and they follow the Dyson-Gorkov equations 
\begin{equation}
  \hat{G}(k)=\hat{G}^{(0)}(k)
 +\hat{G}^{(0)}(k)\hspace{1mm}\hat{\Sigma}^{(1)}(k)\hspace{1mm}\hat{G}(k),
\end{equation}
where $\hat{G}^{(0)}(k)$ is a matrix for
the non-interacting Green's function. 
Within the FLEX approximation, components of the self-energy matrix 
$\Sigma^{(1)}_{ml}(k)$ are given by
\begin{equation}
   \Sigma^{(1)}_{ml}(k)=\frac{T}{N_0}\sum_{q}\sum_{\mu\nu}
    V^{\rm eff}_{\mu m,\nu l}(q)G_{\mu\nu}(k-q),
\end{equation}
with 
\begin{eqnarray}
   &&V^{\rm eff}_{\mu m,\nu l}(q)
   =[\frac{3}{2}\hat{U}^{\rm s}\hat{\chi}^{\rm s}(q)\hat{U}^{\rm s}
    +\frac{1}{2}\hat{U}^{\rm o}\hat{\chi}^{\rm o}(q)\hat{U}^{\rm o}\nonumber\\
   &&\hspace*{5.0mm}-\frac{1}{4}(\hat{U}^{\rm s}+\hat{U}^{\rm o})
     \hat{\overline{\chi}}(q)(\hat{U}^{\rm s}+\hat{U}^{\rm o})
    +\frac{3}{2}\hat{U}^{\rm s}-\frac{1}{2}\hat{U}^{\rm o}]_{\mu m,\nu l},
\end{eqnarray}
where the first and second terms of $V^{\rm eff}_{\mu m,\nu l}(q)$
give contributions of 
the spin and orbital fluctuations, respectively, to the self-energy. 

In Eq.~(4), $\hat{\chi}^{\rm s}(q)$ and $\hat{\chi}^{\rm o}(q)$ are 
the 4$\times$4 matrices of the spin and orbital fluctuations, given by 
\begin{eqnarray}
  &&\hat{\chi}^{\rm s}(q)=
  [\hat{1}-\hat{U}^{\rm s}\hat{\overline{\chi}}(q)]^{-1}
  \hat{\overline{\chi}}(q),\\
  &&\hat{\chi}^{\rm o}(q)=
  [\hat{1}+\hat{U}^{\rm o}\hat{\overline{\chi}}(q)]^{-1}
  \hat{\overline{\chi}}(q),
\end{eqnarray}
where the matrix element of the irreducible 
susceptibility $\overline{\chi}_{ij,st}(q)$ is given by
\begin{equation}
  \overline{\chi}_{ij,st}(q)
   =-\frac{T}{N_0}\sum_{k}G_{si}(k+q)G_{jt}(k),
\end{equation}
For the susceptibility matrices, the labels of row and column 
appear in the order 11, 22, 12, and 21,
pairs of orbital indices 1 and 2.
In these expressions, 
$T$ is a temperature, $N_0$ is number of unit cells, 
abbreviations $k$=$({\bf k},i\omega_{n})$ and 
$q$=$({\bf q},i\Omega_{n})$ 
($\omega_{n}$=$(2n+1)\pi T$, $\Omega_{n}$=$2n\pi T$) are used. 
The interaction matrices 
$\hat{U}^{\rm s}$ and $\hat{U}^{\rm o}$ are given by
\begin{eqnarray}
  \hspace*{-5mm}\hat{U}^{\rm s}
 =\left[\begin{array}{cccc}
   U & 0 & 0 & 0 \\
   0 & U & 0 & 0 \\
   0 & 0 & U' & 0\\
   0 & 0 & 0 & U'
        \end{array}\right],\hspace{5mm}
    \hat{U}^{\rm o}
 =\left[\begin{array}{cccc}
   U & 2U' & 0 & 0 \\
   2U' & U & 0 & 0 \\
   0 & 0 & -U' & 0\\
   0 & 0 & 0 & -U'
        \end{array}\right].\nonumber
\end{eqnarray}
By solving the self-consistent equations, 
the spin and orbital fluctuations and the Green's functions 
for $f$-electrons are determined simultaneously. 

In order to discuss superconductivity, it is necessary to calculate
anomalous self-energy.
The matrix element of the anomalous self-energy 
$\Sigma^{(2)}_{ml}(k)$ is obtained by the functional derivative 
of the thermodynamic potential with respect to 
the anomalous Green's function as follows
\begin{equation}
    \Sigma^{(2)}_{ml}(k)=\frac{T}{N_0}\sum_{q}\sum_{\mu\nu}
    V^{\xi}_{\mu m,l\nu}(q)F_{\mu\nu}(k-q),
\end{equation}
with 
\begin{equation}
    \hat{F}(k)=\hat{G}(k)\hat{\Sigma}^{(2)}(k)\hat{G}(-k),
\end{equation}
where the matrix elements of the effective pairing interactions 
for spin-singlet and spin-triplet channels are given, respectively, by
\begin{eqnarray}
   &&V^{\rm S}_{\mu m,l\nu}(q)
   =[-\frac{3}{2}\hat{U}^{\rm s}\hat{\chi}^{\rm s}(q)\hat{U}^{\rm s}
    +\frac{1}{2}\hat{U}^{\rm o}\hat{\chi}^{\rm o}(q)\hat{U}^{\rm o}\nonumber\\
   &&\hspace{20mm}
     -\frac{1}{2}(\hat{U}^{\rm s}+\hat{U}^{\rm o})]_{\mu m,l\nu},\\
   &&V^{\rm T}_{\mu m,l\nu}(q)
   =[\frac{1}{2}\hat{U}^{\rm s}\hat{\chi}^{\rm s}(q)\hat{U}^{\rm s}
    +\frac{1}{2}\hat{U}^{\rm o}\hat{\chi}^{\rm o}(q)\hat{U}^{\rm o}\nonumber\\
   &&\hspace{17mm}
     -\frac{1}{2}(\hat{U}^{\rm s}+\hat{U}^{\rm o})]_{\mu m,l\nu}.
\end{eqnarray}
The $T_{\rm c}$ is obtained by the temperature at which the maximum
eigenvalue of Eq.~(8) becomes unity. 
As already pointed out in the previous works based on the RPA
\cite{Takimoto,manybody},
one can see from these effective pairing interactions 
that developments of both spin and orbital 
fluctuations have a destructive interference for 
the singlet channel, while they are constructive for 
the triplet one.

%%%%%%%%%%%%%%%%%%% Calculated Results %%%%%%%%%%%%%%%%%%%%%

\section{Calculated results}

The FLEX calculation is numerically carried out for each value
of $\varepsilon$ at fixed parameter values of
$U$=$U'$=4 and $n$=1 corresponding to one $f$-electron density per site.
All summations involved in the above self-consistent equations are
performed using the fast Fourier transformation algorithm
for the ${\bf k}$-space with $32\times32$ meshes in the first Brillouin zone 
and for Matsubara frequency sum with energy cut-off
five times larger than the relevant band width.
%In the following, we present the results obtained by the numerical 
%calculations. 
In Fig.~1, ${\bf q}$ dependences of the principal components of 
$\hat{\chi}^{\rm s}({\bf q},0)$ and
$\hat{\chi}^{\rm o}({\bf q},0)$ are shown for a fixed temperature
$T$=0.02 for different level splitting, $\varepsilon$:
The upper panel for $\varepsilon$=0 and the lower panel for $\varepsilon$=2. 
For $\varepsilon$=0,
corresponding to the orbitally degenerate case,
the AFM spin fluctuation in the $\tau$=1 orbital is enhanced,
but not sufficiently developed to induce $d_{x^2-y^2}$-wave
superconductivity.
With increasing the orbital splitting energy to $\varepsilon$=2,
the AFM spin fluctuation for the $\tau$=1 orbital further develops,
and orbital fluctuations are completely suppressed,
in comparison with the developed AFM spin fluctuation.

%%%%%%%%%%%%%%%%%%%%%%%%%% FIG.1 %%%%%%%%%%%%%%%%%%%%%%%%%%%%%%%%%%%%%
\begin{figure}[t]
\centerline{\epsfxsize=8.5truecm \epsfbox{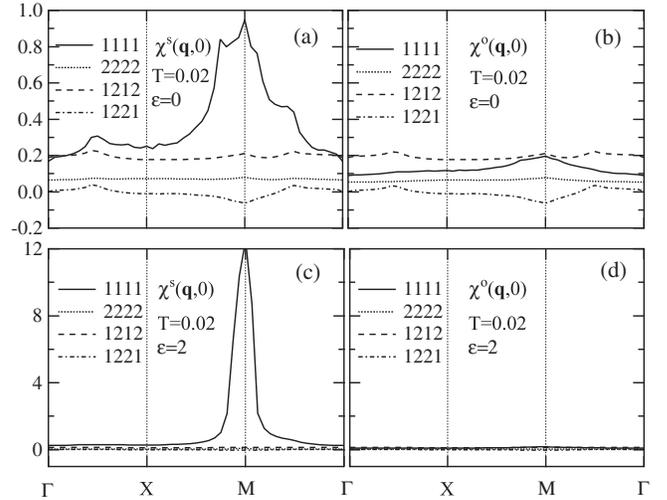} }
%\label{fig1}
\caption{(a) Spin  and (b) orbital susceptibilities in {\bf q} space
for $\varepsilon$=0. 
(c) and (d) are for $\varepsilon$=2.}
\end{figure}
%%%%%%%%%%%%%%%%%%%%%%%%%%%%%%%%%%%%%%%%%%%%%%%%%%%%%%%%%%%%%%%%%%%%%%%

In Fig.~2, the phase diagram obtained within the FLEX approximation
for the orbital degenerate system are shown, where the solid and open
circles describe the superconducting and AFM transitions, respectively.
The dotted curve is a schematic phase boundary expected
between the two phases. From Fig.~2, we can see that
(1) the spin-singlet superconducting phase with $B_{\rm 1g}$-symmetry
appears next to the AFM phase and (2) $T_{\rm c}$ is enhanced with
increasing the orbital splitting energy $\varepsilon$.
From these observations, we can conclude that the superconducting phase
is induced by the development of the AFM spin fluctuations
for the $\tau$=1 orbital with increasing the orbital splitting energy 
$\varepsilon$.

%%%%%%%%%%%%%%%%%%%%%%%%%% FIG.2 %%%%%%%%%%%%%%%%%%%%%%%%%%%%%%%%%%%%%
\begin{figure}[t]
\centerline{\epsfxsize=8.5truecm \epsfbox{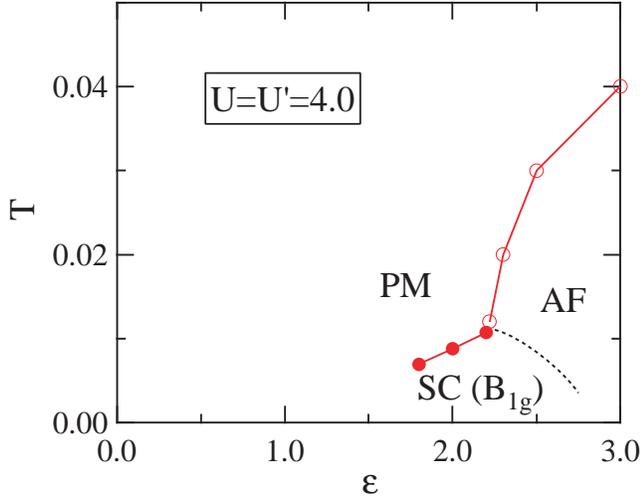} }
%\label{fig2}
\caption{
Phase diagram in the $T$-$\varepsilon$ plane for $U$=$U'$=4.0 
obtained by the FLEX approximation.}
\end{figure}
%%%%%%%%%%%%%%%%%%%%%%%%%%%%%%%%%%%%%%%%%%%%%%%%%%%%%%%%%%%%%%%%%%%%%%%

%%%%%%%%%%%%%%%%%%%%%  Summary  %%%%%%%%%%%%%%%%%%%%%%%%%

\section{Summary and discussion}

Let us now discuss the experimental results for CeTIn$_5$ in comparison
with the present theoretical results.
First we should note that
in the actual tetragonal crystal, the four-fold degenerate $\Gamma_8$
states in the cubic notation split into two Kramers doublets,
$\Gamma_7$ and $\Gamma_6$, as we mentioned in the Sec.~2.
Therefore, in the tetragonal system, the $j$=5/2 states split into
two $\Gamma_7$ and one $\Gamma_6$.
Analyses of experimental data of magnetic susceptibility of
CeTIn$_5$ by using the CEF theory seem to be consistent more
with the level scheme where two $\Gamma_7$ are lower than the
$\Gamma_6$ \cite{Takeuchi}.
The energy splitting between the two $\Gamma_7$ is estimated
as 68K for CeRhIn$_5$, 61K for CeIrIn$_5$, and 151K for CeCoIn$_5$.
According to the present analysis, higher $T_{\rm c}$ is obtained
for larger $\varepsilon$,
consistent with the tendency in $T_{\rm c}$ of
the two superconducting materials: $T_{\rm c}$=0.4K for CeIrIn$_5$
and $T_{\rm c}$=2.3K for CeCoIn$_5$.

However, the antiferromagnetically ordered CeRhIn$_5$ with $T_{\rm N}$=3.8K
has an intermediate value for $\varepsilon$.
One possible scenario to understand the discrepancy is to consider
difference of quasi-two dimensionality, as we have pointed out
in \cite{manybody}.
To make more direct and quantitative comparison with experimental
results, however, one should be aware of an assumption for
the model used in the present study,
which may be called as $\Gamma_8$ model.
Namely, this $\Gamma_8$ model assumes that $\Gamma_7$ and $\Gamma_6$
are the lower two Kramers doublets, which may be different from
the level scheme obtained from the experiments.
Thus, it will be an interesting future problem to elucidate the role
of orbital splitting for the more realistic two $\Gamma_7$ model,
which includes not only the nearest neighbor hopping but also
the next nearest neighbor one and thus, may reproduce the realistic
electronic states better than the $\Gamma_8$ model \cite{manybody2}.

In order to construct a whole story for CeTIn$_5$ compounds,
it may be required to use the even more realistic $f$-$p$ model
including the $f$-$p$ hybridization explicitly.
In the $f$-$p$ model, we can have two different energy scales:
The large energy scale is the band width for the conduction electron
and/or the Coulomb interaction,
while the small energy scale corresponds to the energy splitting
of the CEF levels discussed here.
The present study indicates the possibility that the small energy scale
of the CEF level splitting plays a key role as the controlling parameter
of quantum phase transitions.

In summary, based on the effective microscopic model with
orbital degeneracy for $f$-electron systems,
we have proposed that the orbital splitting energy is
the parameter controlling the change from the paramagnetic
to the AFM phase with the $d_{x^2-y^2}$-wave superconducting
phase in between.
Actually, the self-consistent FLEX approximation applied
to the orbital degenerate model shows that the $d_{x^2-y^2}$-wave 
superconducting phase is induced by increasing the orbital 
splitting energy.

%%%%%%%%%%%%%%%%%%%%% Acknowledgement %%%%%%%%%%%%%%%%%%%%%%%%%%%%%

\section*{Acknowledgement}

The authors would like to thank T. Maehira for many helpful discussions.
T.H. and K.U. are supported by the Grant-in-Aid for Scientific Research 
from Japan Society for the Promotion of Science.

\vskip-0.5cm
%%%%%%%%%%%%%%%%%%%%% References %%%%%%%%%%%%%%%%%%%%%%%%%%%%%%%

\end{document}